%% file: main.tex
\documentclass[a4paper,aps,prd,showpacs,twocolumn, nofootinbib
]{revtex4-1}
\usepackage{preamble}
\usepackage{hyperref}

\usepackage{graphicx}

\begin{document}
\title{Gauge invariant quantum backreaction in $U(1)$ axion inflation}

\author{Davide Campanella Galanti$^{a,b}$}
\email{d.campanellagalan@studenti.unipi.it}
\author{Pietro Conzinu$^{c,d,b}$}
\email{pietro.conzinu@cern.ch}
\author{Giovanni Marozzi$^{a,b}$}
\email{giovanni.marozzi@unipi.it}
\author{Simony Santos da Costa$^{e,f,b}$}
\email{simony.santosdacosta@unitn.it}
\affiliation{$^{a}$Dipartimento di Fisica, Universit\`a di Pisa, Largo B. Pontecorvo 3, 56127 Pisa, 
Italy\\
$^{b}$Istituto Nazionale di Fisica Nucleare, Sezione di Pisa, Italy
\\
$^{c}$Dipartimento di Scienze Matematiche, Fisiche e Informatiche, Università di Parma, Parco Area delle Scienze $7/A, I-43124$, Parma, Italy\\
$^{d}$Theory Division, CERN, CH-1211 Geneva 23, Switzerland
\\
$^{e}$Dipartimento di Fisica, Universit\`a di Trento, Via Sommarive 14, 38123 Trento, Italy\\
$^{f}$Trento Institute for Fundamental Physics and Applications (TIFPA), Via Sommarive 14, 38123 Trento, 
Italy
}

\begin{abstract}
We evaluate the quantum backreaction due to a gauge field coupled to a pseudo-scalar field driving a slow-roll inflationary stage, the so-called axion inflation.
The backreaction is evaluated for the first time using a gauge invariant approach, going to second order in perturbation theory, and taking into consideration inflaton fluctuations as well as scalar perturbations of the metric. 
Within our gauge-invariant, but observer-dependent approach, we naturally consider as physical observers the ones comoving with the inflaton field.
Considering the effective expansion rate consequent to the gauge field's backreaction, we observe that the backreaction effect becomes significant quite rapidly, moving the system out of the perturbative regime and into what is often referred to as the strong backreaction regime. This behavior also applies to the parameter that dictates the production of the gauge fields.  
The space-time backreaction is mainly due to the helicity contribution within the region of validity of the perturbative regime.
As a final result, we see that the evaluated backreaction goes in the direction of prolonging the inflationary period more compared to the scenarios previously studied.
\end{abstract}

\maketitle

\section{Introduction}
\label{Int}

The inflationary paradigm was originally proposed to solve the problems associated with the Hot Big-Bang standard theory \cite{Guth:1980zm,Linde:1981mu} via an early stage of accelerated cosmological expansion. Immediately afterwards it was also realized that already a single field inflationary model could furnish a causal mechanism to generate nearly scale invariant spectra of scalar and tensor fluctuations \cite{Mukhanov:1981xt,Starobinsky:1979ty}, elevating the model to one of the most recognized and accepted candidates to describe the early universe.

This first success leads to a continuous and intense investigation of different inflationary models, together with the investigation of the fundamental nature of the inflaton field, its possible interaction with other fields, and the predictions and phenomenology associated to the possible different models (see for example \cite{Martin:2013tda} for a large collection of models).
Recent literature \cite{Anber:2009ua, Sorbo:2011rz, Anber:2012du, Barnaby:2011vw} has generated significant interest in the behavior of a pseudo-scalar inflaton when coupled to \textit{U}(1) \textit{abelian} gauge fields.
This possibility extends models of inflation driven by an axion-like particle, such as \textit{natural inflation}, first proposed in \cite{Freese:1990rb, Adams:1992bn}. 
Axions appear copiously in string theory \cite{Svrcek:2006yi, Kim:1986ax, Marsh:2015xka} providing a notable inflaton candidate in UV complete theory.

Within this framework, the inflaton/gauge field coupling brings several interesting effects with a rich phenomenology (see, for example, 
\cite{Ballardini:2019rqh, Anber:2009ua,Pajer:2013fsa, Maleknejad:2012fw, Peloso:2022ovc, Barnaby:2011qe, Turner:1987bw,Garretson:1992vt,Adshead:2016iae,Sobol:2019xls,Adshead:2015pva,McDonough:2016xvu,Domcke:2018eki, Durrer:2023rhc,Barnaby:2011vw,Domcke:2020zez, Caravano:2021bfn, Hashiba:2021gmn, Ishiwata:2021yne, Lozanov:2018kpk}). 
Furthermore, let us mention how inflation driven by a pseudo-scalar is the perfect touchstone to incorporate parity violation during a quasi-exponential expansion (see e.g. \cite{Bartolo:2014hwa}).

The core of the physical process concerning the coupling between the pseudo-scalar inflaton $\phi$ and the gauge fields $A_{\mu}$ relies on the so-called Chern-Simons-like operator.
As a consequence of the coupling through such operator, quantum vacuum fluctuations of the gauge field $A_{\mu}$ are amplified into physical excitations, since $A_{\mu}$ acquires a time-dependent background as a consequence of the slow-rolling of the inflaton. In this way, the field $\phi$ will slow down because its energy is utilized for the production of the quantum gauge fields: this dissipative effect should make inflation lasts longer depending on how much the 
gauge coupling is strong \cite{Anber:2009ua, Sorbo:2011rz, Anber:2012du, Barnaby:2011vw}. 
To resume, as a consequence of the pseudo-scalar/gauge fields coupling, the inflaton field provokes an amplification of the quantum fluctuations of $A_{\mu}$ which backreact on the space-time dynamic.
\\

Let us mention that, since we are in the framework of QFT in curved space-time the observables present in the model 
show (as usual) ultraviolet (UV) divergences and have to be regularized through a renormalization scheme \cite{Birrell:1982ix, Parker:2009uva}. Here we will use the adiabatic regularization approach \cite{Zeldovich:1971mw, Parker:1974qw, Fulling:1974zr, Bunch:1980vc, Anderson:1987yt} in order to compute the energy density and helicity integral of the gauge fields. In particular, we will consider in our computation the finite results obtained through a new adiabatic renormalization scheme which use a physically motivated comoving infrared (IR) cut-off, as described in \cite{Animali:2022lig}, in order to remove some unphysical IR divergences that one would obtain through the standard adiabatic subtraction for the considered model (see \cite{Ballardini:2019rqh}).

In this manuscript we evaluate the gauge fields backreaction 
using a gauge invariant observer dependent averaging prescription \cite{Gasperini:2009wp} and 
looking to a set of gauge invariant effective field equations as defined in \cite{Gasperini:2009mu} (see also the seminal works \cite{Buchert:1999er,Buchert:2001sa}). 
Furthermore, following \cite{Finelli:2011cw} we will work within a perturbative framework and go up to the  
second perturbative order in the evaluation of the cosmological backreaction to obtain the first non-zero contribution.  

The set of effective equations of \cite{Gasperini:2009mu} describes, in particular, the evolution of the effective expansion rate of the universe as measured by a given observer. Within the framework of single field inflation is then natural to consider as such observer the one comoving with the inflaton field, being the inflaton field the only available clock in  this framework. 
In this approach, where one looks to cosmological observables, Einstein equations are satisfied at any perturbative order (as one would expect) and the gauge fields backreaction on the expansion rate is a consequence of their coupling with the matter and metric scalar fluctuations to second perturbative order. 
The above approach differs from the standard one proposed in \cite{Anber:2009ua, Anber:2012du, Barnaby:2011qe}, and further developed in \cite{Gorbar:2021rlt, Ballardini:2019rqh, Adshead:2016iae,Sobol:2019xls,Adshead:2015pva,McDonough:2016xvu,Domcke:2018eki,Domcke:2020zez, Hashiba:2021gmn, Ishiwata:2021yne, Lozanov:2018kpk}. In the case of \cite{Anber:2009ua, Anber:2012du, Barnaby:2011qe} the backreaction of the gauge fields is studied adding their contribution to the background Friedmann equations (with Einstein equation effectively considered only at the background level) and to the background equation of motion of the pseudo scalar inflaton field.
Namely, in \cite{Anber:2009ua, Anber:2012du, Barnaby:2011qe} is taken in consideration only the gauge fields contribution, but are neglected the induced scalar perturbations of matter and metric sectors, although the inflaton fluctuations were then added in \cite{Domcke:2023tnn} and simulations in lattice were provided in \cite{Caravano:2021bfn, Caravano:2022yyv, Caravano:2022epk, Cheng:2015oqa,  Figueroa:2023oxc, Figueroa:2021yhd}.

The manuscript is organized as follows. 
In Section \ref{Sec2} we describe the theoretical framework of the U(1)-axion inflation model, where a pseudo-scalar inflaton is coupled to abelian gauge fields. 
In Section \ref{Sec3} we introduce the gauge invariant, observed dependent, approach to the cosmological backreaction problem we shall use, and explicitly obtain the related effective Hubble expansion rate. 
In Section \ref{Sec4} we detail the second-order perturbative equations for scalar perturbations taking into consideration their coupling with the gauge field fluctuations.
In Section \ref{Sec5} we introduce the class of the comoving observers in single-field inflationary models and describe the form of the backreaction terms coming from the gauge fields contribution. 
In Section \ref{Sec6} we solve the second-order equation of motion for the pseudo-scalar inflaton, considering the gauge fields contribution, and present the analytical expression for the gauge fields backreaction on the effective expansion rate. 
In Section \ref{sec7} we introduce the numerical methods used to evaluate the magnitude of the gauge invariant backreaction effect, and plot the consequent results. 
Finally, in Section \ref{Conc} we present our final remarks and conclusions.
In Appendix \ref{app.A} we give the expressions for the energy density and the helicity integral of the gauge fields used in the manuscript while in Appendix \ref{app.B} we write the second order perturbative expression for Einstein tensor and energy-momentum tensor.

\section{$U(1)$-axion inflation model}
\label{Sec2}

Let us consider a pseudo-scalar inflaton coupled to \textit{U}(1) \textit{abelian} gauge fields described by the following action 
{\fontsize{3mm}{3mm}
\begin{align}\label{eq:4.1-3}
    &S=\int\dd^4 x\sqrt{-g}\le[ -\dfrac{1}{2}\partial_\mu \phi \partial^\mu \phi-V(\phi)-\dfrac{1}{4}F_{\mu \nu}F^{\mu \nu}-\dfrac{g\phi}{4}F^{\mu \nu}\tilde{F}_{\mu\nu} \ri]\,,
\end{align}}
where $\phi$ is the axion-like field,  $\tilde{F}^{\mu\nu}=\frac{1}{2} \epsilon^{\mu\nu \alpha \beta}F_{\alpha \beta}$, $F_{\alpha \beta}=\partial_{\alpha}A_{\beta}-\partial_{\beta}A_{\alpha}$,  $A_\mu$ is the gauge fields. Furthermore, the coupling constant $g$ can be expressed in terms of the axion decay constant $f$ by the relation $g = \alpha/f$, with $\alpha$ a dimensionless parameter. The last term in the action \eqref{eq:4.1-3} is a Chern-Simon like term, thanks to this coupling term the axion field drives the gauge fields amplification.

To make contact with past literature, let us first consider the case in which we set to zero metric perturbations. 
Therefore, we consider a spatially flat Friedmann-
Lema\^{ı}tre-Robertson-Walker (FLRW) background metric
\begin{equation}
    ds^2=-dt^2+a^2(t)d\textbf{x}^2 \,,
\end{equation}
with $a(t)$ the scale factor,
and we will evaluate the impact of the gauge fields production on the background Euler-Lagrange and Einstein equations.
Starting from the Euler-Lagrange equations, using the {\it{electro-magnetic-}}notation 
(bold letters stand for 3 dimensional vectorial quantities)
\begin{align}
    &\textbf{E}=-\dfrac{\textbf{A}'}{a^2} \,, \qquad \qquad  \textbf{B}=\frac{\pmb{\nabla}\times \textbf{A}}{a^2} \,,
\end{align}

one finds \cite{Anber:2009ua}
\begin{align}
    &\phi''+2aH \phi'-\laplacian{\phi}+a^2 V_{\phi}(\phi)=g a^2 \textbf{E} \cdot \textbf{B} \,, \label{EqPhiNoAv}\\ \nonumber \\
    &\textbf{E}'+2aH\textbf{E}-\pmb{\nabla} \times \textbf{B}=-g \phi' \textbf{B}-g \pmb{\nabla} \phi \times \textbf{E} \,,\\ \nonumber \\
    &\pmb{\nabla} \cdot \textbf{E}=-g \qty(\pmb{\nabla} \phi)\cdot \textbf{B} \,,
\end{align}
while from the Bianchi identities one obtains
\begin{align}
    &\textbf{B}'+2aH\textbf{B}+\pmb{\nabla} \times \textbf{E}=0 \,,\\ \nonumber \\
    & \pmb{\nabla} \cdot \textbf{B}=0 \,,
\end{align}
where the prime indicates the derivative with respect to the conformal time and $V_{\phi}=\dv{V}{\phi}$. 

On studying only the impact of the gauge fields on the inflationary background, we take the inflaton as homogeneous, $\pmb{\nabla}\phi=0$.
As a consequence, in the Coulomb gauge $(A_0=0\,,\;\pmb{\nabla} \cdot {\bf{A}}=0)$, we obtain the following equation of motion for the gauge field
\begin{equation}
    \pdv[2]{\textbf{A}}{\tau}-\laplacian{\textbf{A}}-g\phi'\pmb{\nabla}\times {\textbf{A}}=0 \,.
\end{equation}

We then promote $\textbf{A}(\tau, \textbf{x})$ to a quantum operator
\begin{equation}
    \hat{\textbf{A}}(\tau, \textbf{x})=\sum_{\lambda=\pm} \bigintsss \dfrac{d^3k}{(2\pi)^3} \qty[\pmb{\epsilon}_{\lambda}(\textbf{k})A_{\lambda}(\tau,\textbf{k})\hat{a}_{\lambda}(\textbf{k})e^{i \textbf{k}\textbf{x}}+\text{h.c.}] \,,
\end{equation}
where we denote with $\pmb{\epsilon}_{\pm}$ the polarization vectors of the gauge field, and choose as a basis the circular polarization transverse to the direction of propagation fixed by the (comoving) momentum $\textbf{k}$.
The helicity vectors satisfy then the following relations
\beq
\begin{matrix}
&\textbf{k} \cdot \pmb{\epsilon}_{\pm}(\textbf{k})=0 \,, && \quad\textbf{k} \times \pmb{\epsilon}_{\pm}(\textbf{k})=\mp i \abs{\textbf{k}} \pmb{\epsilon}_{\pm}(\textbf{k}) \,,\\
\\
    &\pmb{\epsilon}^{\ast}_{\pm}(\textbf{k})=\pmb{\epsilon}_{\pm}(-\textbf{k}) \,, 
    &&\pmb{\epsilon}^{\ast}_{\lambda}(\textbf{k})\cdot \pmb{\epsilon}_{\sigma}(\textbf{k})=\delta_{\lambda \sigma} \,.
\end{matrix}
\eeq

Finally, the equation of motion for the Fourier modes $A_{\pm}$ reads
\begin{equation}\label{eq:4.1-6}
    \dv[2]{}{\tau}A_{\pm}(\tau, k)+\qty(k^2\mp k g \phi')A_{\pm}(\tau, k)=0 \,.
\end{equation}
To understand the impact of the gauge fields on the Einstein equations one has to evaluate their energy-momentum tensor.
Starting from the initial action we then obtain the following result
\begin{equation}\label{eq:4.1-15}
    T_{\mu \nu}^{(F)}=F_{\rho \mu}F^{\rho}_{\hspace{0.1cm}\nu}+g_{\mu \nu} \dfrac{\textbf{E}^2-\textbf{B}^2}{2} \,,
\end{equation}
where the energy density and pressure terms, descend from the $(0, 0)$ and $(i, j)$ components of $T_{\mu \nu}^{(F)}$. These are given by
\begin{align}
    T_{00}^{(F)}&=\dfrac{\textbf{E}^2+\textbf{B}^2}{2} \,, \\ 
    T_{ij}^{(F)}&=-E_i E_j- B_i B_j + \delta_{ij} \dfrac{\textbf{E}^2+\textbf{B}^2}{2} 
    \label{TijF}\,.
\end{align}
Let us also note that the off-diagonal term $T^{0\hspace{0.05cm}(F)}_{\hspace{0.1cm}i}=F_{j}^{\hspace{0.1cm}0}F^{j}_{\hspace{0.1cm}i}$ can be rewritten as
\begin{equation}
    T^{0\hspace{0.05cm}(F)}_{\hspace{0.1cm}i}=F_{j}^{\hspace{0.1cm}0}F^{j}_{\hspace{0.1cm}i}=\epsilon_{ijk}E^jB^k=\qty(\textbf{E}\times \textbf{B})_i \,,
\end{equation}
where $\epsilon_{ijk}$ is the Levi-Civita tensor.

Taking into account only the backreaction of the gauge fields on the background, we can then rewrite the Friedmann equations in the Hartree approximation as (see e.g. \cite{Anber:2009ua}) 
\begin{align}
    &H^2=\dfrac{1}{3 M^2_{Pl}}\qty[\dfrac{\dot{\phi}^2}{2}+V(\phi)+\dfrac{\langle \textbf{E}^2+\textbf{B}^2 \rangle}{2}] \label{eq:H^2standard}\,,\\ \nonumber\\
    &\dot{H}=-\dfrac{1}{2 M^2_{Pl}}\qty[\dot{\phi}^2+\dfrac{2}{3}\langle \textbf{E}^2+\textbf{B}^2 \rangle] \,,
    \label{HBRV}
\end{align}
where we have used $M^2_{Pl}=\frac{1}{8\pi G}$. While, the background inflaton equation of motion correspondent to 
Eq.(\ref{EqPhiNoAv}) becomes (in proper time): 
\begin{equation}\label{eq:4.1-11}
    \ddot{\phi}+3H\dot{\phi}+V_{\phi}=g \langle \textbf{E} \cdot \textbf{B} \rangle \,.
\end{equation}
with $\VEV{\cdots}$ the (quantum) expectation value. The exact expression for the energy density $\Ene$ and the so-called helicity integral $\Hel$ that we will use in the following to evaluate analytically and numerically the backreaction are the ones obtained in \cite{Animali:2022lig}. For the sake of completeness and clarity we report them in Appendix \ref{app.A}.

\section{Gauge invariant quantum backreaction}
\label{Sec3}

Cosmological backreaction is a consequence of the non-linearity of the Einstein equations. 
Given a background cosmological solution of metric and matter, the addition of small amplitude fluctuations leads to perturbative equations which depend on the perturbative order considered. 
As an example, to the quadratic order we obtain a quadratic first order source for our second order perturbations, and so on at higher perturbative order. 
An observer in a homogeneous universe would measure different physical features from those perceived by dynamical observers living in a perturbed space-time. This effect is indeed what we call the cosmological backreaction. 
The cosmological backreaction induced by quantum fluctuations along the inflationary epoch have been, in particular, the subject of several investigations relying on many different approaches \cite{Mukhanov:1996ak, Abramo:1997hu, Unruh:1998ic, Abramo:1998hj, Abramo:2001dc, Abramo:2001db, Geshnizjani:2002wp, Losic:2005vg, Finelli:2003bp}. 
In particular, starting from the observation that a backreaction effect is meaningful only when ``measured'' by a given observer, a covariant and gauge invariant (GI) approach was constructed, following \cite{Gasperini:2009wp, Gasperini:2009mu}, within the framework of quantum backreaction in the early universe in \cite{Finelli:2011cw}. This approach was then further applied to different cases within the framework of slow-roll inflation (see \cite{Brandenberger:2018fdd, Marozzi:2014xma, Marozzi:2013uva, Marozzi:2012tp, Marozzi:2011zb}).
In practice, by using gauge invariant (but observer dependent) averaged quantities, as defined in \cite{Gasperini:2009wp}, one can  describe the space-time evolution by a set of gauge invariant effective observer-dependent equations \cite{Gasperini:2009mu}.
Going into more details, one can construct a general non-local observable by taking quantum averages of a scalar field $S(x)$ over a space-time hyper-surface 
$\Sigma_{A_0}$ where $A(x)$, another time-like scalar field, assumes a constant value $A_0$. 
By setting a barred coordinate system $\bar{x}^\mu=(\bar{t}, \textbf{x})$, in which the scalar field $A$ is homogeneous,
the above physical gauge invariant observable is then defined by \cite{Gasperini:2009wp, Gasperini:2009mu}
\begin{equation}\label{eq:media}
    \langle S \rangle_{A_0}=\dfrac{\langle \sqrt{\abs{\overline{\gamma}(t_0, \textbf{x})}} \hspace{0.1cm}\overline{S}(t_0, \textbf{x}) \rangle}{\langle \sqrt{\abs{\overline{\gamma}(t_0, \textbf{x})}}\rangle} \,,
\end{equation}
where $t_0$ is the time $\bar{t}$ for which $\bar{A}(\bar{x})=A^{(0)}(\bar{t})=A_0$, and $\abs{\overline{\gamma}(t_0, \textbf{x})}$ is the determinant of the induced three dimensional metric on $\Sigma_{A_0}$.

The four vector
\begin{equation}
    n^{\mu}=-\dfrac{\partial^{\mu}A}{\sqrt{-\partial^{\nu}A\partial_{\nu}A}} \,,
\end{equation}
settles then the class of observers and the \textit{foliation} of the space-time.

Let us now look to the effective scale factor $a_\eff=\langle \sqrt{\abs{\overline{\gamma}}}\rangle ^{\frac{1}{3}}$, which describes the expansion of the space-time as seen by the observers comoving with the hyper-surface $\Sigma_{A_0}$. This effective scale factor satisfies the following gauge independent effective equation \cite{Gasperini:2009mu}
\begin{equation}\label{eq:5.1.1-1}
    \qty(\dfrac{1}{a_\eff}\pdv{a_\eff}{A_0})^2=\dfrac{1}{9}\biggl\langle \dfrac{\Theta}{\sqrt{-\partial^{\mu}A\partial_{\mu}A}}\biggl\rangle^2_{A_0} \,,
\end{equation}
where $\Theta=\nabla_{\mu}n^{\mu}$ is the expansion scalar of the time-like congruence $n^{\mu}$. As a consequence of the above gauge invariant construction, the space-time dynamics  can be studied in any gauge, simply by solving the matter and Einstein equations in the chosen gauge. 

Following \cite{Finelli:2011cw}, to evaluate the backreaction within a perturbative framework
we consider cosmological perturbation theory up to the second order,
so we write the general perturbed metric around a FLRW  space-time as
\begin{align}
    g_{00}&=-1-2\alpha-2\alpha^{(2)} \,,\nonumber \\
    g_{i0}&=-\dfrac{a}{2}\qty(\beta_{,\hspace{0.05cm}i}+B_i)-\dfrac{a}{2}\qty(\beta^{(2)}_{,\hspace{0.05cm}i}+B^{(2)}_i) \,,\nonumber \\
    g_{ij}&=a^2 \biggl[\delta_{ij}\qty(1-2\psi-2\psi^{(2)})+D_{ij}\qty(E+E^{(2)})\nonumber \\&+\dfrac{1}{2}\qty(\chi_{i,j}+\chi_{j,i}+h_{ij})+\dfrac{1}{2}\qty(\chi^{(2)}_{i,j}+\chi^{(2)}_{j,i}+h^{(2)}_{ij})\biggr] \,,
    \label{GeneralGauge}
\end{align}
where $D_{i j}=\partial_i \partial_j -\delta_{i j} \nabla^2/3$ and, for clearness, we write first order quantities without an upper script.
We then have that $\alpha$, $\beta$, $\psi$ and $E$ are scalar perturbations, $B_i$ and $\chi_i$ are 
transverse vectors ($\partial^i B_i=0$ and $\partial^i \chi_i=0$), and $h_{i j}$ is a traceless and transverse tensor ($\partial^i h_{i j}=0$ and $h^i_i=0$). In the follow we also neglect vector perturbations which decay dynamically.
 
At the same time, the inflaton can be written to second order as
\begin{equation}
    \phi(t,\textbf{x})=\phi(t)+\varphi(t,\textbf{x})+\varphi^{(2)}(t,\textbf{x}) \,.
\end{equation}
Let us note how the ten degrees of freedom in the metric (\ref{GeneralGauge}) are redundant. To fix a gauge corresponds to reduce such degrees of freedom,
typically removing two scalars and one vector perturbations. 
The cases of interests for us are
the uniform field gauge (UFG), defined by setting $\phi(t,\textbf{x})=\phi(t)$ and 
by another conditions (we consider $g_{i0}=0$), and 
the uniform curvature gauge (UCG), defined by
$g_{ij}=a^2\left[\delta_{ij}+\frac{1}{2} \left(h_{ij}+h^{(2)}_{ij}\right)\right]$.

Applying the long wavelength (LW) limit for scalar fluctuations~\footnote{In the long wavelength limit one neglects terms like $\sim |\vec{\nabla} \varphi|^2/(a H)^2$ in evaluating cosmological observables. This is a consequence of the fact that in the Fourier space such terms goes like $k^2/(a H)^2$ and are negligible in the infrared, while their ultraviolet tail is almost erased by the renormalization process.} 
one then gets \cite{Finelli:2011cw}
\begin{align}
    &\bar{\Theta}=3H-3H\bar{\alpha} -3\dot{\bar{\psi}}+\dfrac{9}{2}H \bar{\alpha}^2+3\bar{\alpha} \dot{\bar{\psi}} -6\bar{\psi} \dot{\bar{\psi}}\nonumber \\&  -3H\bar{\alpha}^{(2)}-3\dot{\bar{\psi}}^{(2)}-\dfrac{1}{8}h_{ij}\dot{h}^{ij} \label{eq:theta} \,, \\
   &-\partial_{\mu}\bar{A}\partial^{\mu}\bar{A}=1-2\bar{\alpha}+4\bar{\alpha}^2-2\bar{\alpha}^{(2)} \label{eq:partial}\,, \\
    &\sqrt{\abs{\overline{\gamma}}}=a^3 \qty(1-3\bar{\psi}+\dfrac{3}{2}\bar{\psi}^2-\dfrac{1}{16}h^{ij}h_{ij}-3\bar{\psi}^{(2)}) \label{eq:gamma}\,.
\end{align}
Then by substituting Eqs. \eqref{eq:theta}, \eqref{eq:partial} and \eqref{eq:gamma} inside Eq.~\eqref{eq:5.1.1-1}, and using Eq. (\ref{eq:media}), one obtains the general form of the effective expansion for the class of observers defined by the scalar field $A(t, \textbf{x})$. Thus, neglecting tensor perturbations and in the LW limit, we obtain \cite{Finelli:2011cw}
\begin{equation}
\label{5.1.1-2}
    H^2_\eff=\qty(\dfrac{1}{a_\eff}\pdv{a_\eff}{A_0})^2=H^2\qty[1+\dfrac{2}{H}\langle \bar{\psi}\dot{\bar{\psi}}\rangle-\dfrac{2}{H}\langle \dot{\bar{\psi}}^{(2)} \rangle] \,.
\end{equation}
Let us stress that, since the above result is by construction gauge invariant, we can now use the dynamical solutions for inflaton and metric fluctuations in any frame of our choice.

\section{Second order perturbative equations}
\label{Sec4}
Hereafter, we want to evaluate the gauge fields backreaction taking in consideration their coupling with both matter and metric scalar perturbations, and using the gauge invariant approach described above. Therefore, with the aim of considering the weak backreaction regime we promote the gauge fields $A_{\mu}$ to first order perturbations and consistently consider the scalar sector up to second order in perturbation theory. In fact, within the assumption that gauge fields have no background, such first order fields evolve independently from the matter-gravity scalar perturbation sector. The equations of motion that govern the dynamics of $A_{\mu}$ are unchanged, and one has a coupling among the two different sectors only to second order.

To evaluate how the gauge fields backreaction impact on the effective Hubble factor $H_\eff$ of Eq.(\ref{5.1.1-2}) we have so to solve Einstein and Euler-Lagrange equations up to second order taking in consideration the above coupling.  As described in the previous section, the approach used to obtain $H_\eff$ is gauge invariant and we can fix the gauge without lose of generality. 
 
Let us so choose the uniform curvature gauge (UCG), where, as mentioned, we fix $\psi=E=0$ at first and second order in Eq.~\eqref{GeneralGauge}.  
The Einstein tensor terms remain the usual ones (first obtained in \cite{Finelli:2003bp} and reported in Appendix \ref{app.B}), while the energy-momentum tensor is modified by the presence of the gauge fields.
As said, we consider the weak backreaction regime and neglect the background contribution of the gauge fields. As a consequence, we will treat expressions like $\textbf{E}^2$, $\textbf{B}^2$, $\textbf{E}\cdot \textbf{B}$ as second order perturbations. 
The gauge fields contributions to the energy-momentum tensor $T_{\mu\nu}^{(F)}$ can then be obtained using the background metric and coincide with the one evaluated in Eqs. (\ref{eq:4.1-15})$-$(\ref{TijF}). This will act as a second order source for inflaton and metric fluctuations.
Furthermore, the axionic coupling between the pseudo-scalar inflaton and the gauge fields introduces a new term in the Euler-Lagrange equation (see Eq.(\ref{EqPhiNoAv})).

As said, first order scalar perturbations are decoupled from the gauge fields contribution, while from the momentum constraint to second order one finds (the scalar contributions below and in the follow were first evaluated in \cite{Finelli:2003bp})
\begin{equation}\label{eq:5.2.1-2}
\alpha^{(2)} = 4\pi G \dfrac{\dot{\phi}}{H}\varphi^{(2)} + s + s_{F} \,,
\end{equation}
with
\begin{align}
    s&=-4\pi G \epsilon \varphi^2+2\alpha^2-\dfrac{\abs{\pmb{\nabla}\beta}^2}{8}+\dfrac{1}{\laplacian}\biggl[\dfrac{4\pi G}{H}\pmb{\nabla}\cdot \qty(\dot{\varphi}\pmb{\nabla}\varphi)\nonumber \\& +\dfrac{1}{4aH}\qty(\alpha^{,\hspace{0.05cm}kj}\beta_{,\hspace{0.05cm}kj}-\alpha^{,\hspace{0.05cm}k}_{,\hspace{0.05cm}k}\beta^{,\hspace{0.05cm}j}_{,\hspace{0.05cm}j})\biggr]\,,\\
    s_{F}&=-\dfrac{4\pi G}{H}\dfrac{1}{\laplacian}\qty[\pmb{\nabla}\cdot\qty(\textbf{E}\times \textbf{B})] \,.
\end{align}
From the equivalence between the curvature part and the matter sector of the Einstein equations to second order, we obtain
\begin{align}
    \dfrac{H}{a}\laplacian{\beta^{(2)}}  &= 8 \pi G \dfrac{\dot{\phi}^2}{H}\dv{t}\qty(\dfrac{H}{\dot{\phi}}\varphi^{(2)}) -\qty(Q+Q_{F})  
    \nonumber\\
    &+16 \pi G V (s + s_{F}) \,,
\end{align}
where
\begin{align}
    Q&=12H^2\alpha^2-\dfrac{3}{4}H^2\abs{\pmb{\nabla}\beta}^2-\dfrac{H}{a}\qty(\alpha_{,\hspace{0.05cm}i}\beta^{,\hspace{0.05cm}i}+2\alpha \laplacian{\beta})\nonumber \\&-\dfrac{1}{8a^2}\qty(\beta_{,\hspace{0.05cm}ij}\beta^{,\hspace{0.05cm}ij}-\qty(\laplacian{\beta})^2)-8\pi G \Biggl[\dfrac{\dot{\varphi}^2}{2}+\dfrac{\abs{\pmb{\nabla}\varphi}^2}{2a^2}\nonumber \\&+\dfrac{1}{2}V_{\phi\phi}\varphi^2+2\dot{\phi}^2\alpha^2-2\dot{\phi}\alpha\dot{\varphi}-\dfrac{1}{8}\dot{\phi}^2\abs{\pmb{\nabla}\beta}^2\Biggr] \,, \\
    Q_{F} &=-8\pi G \qty(\dfrac{\textbf{E}^2+\textbf{B}^2}{2}) \,.
\end{align}
On the other hand, the equation of motion for the second order inflaton perturbation, due to the presence of the gauge fields, reads
\begin{align}\label{eq:5.2.1-1}
    &\ddot{\varphi}^{(2)}+3H\dot{\varphi}^{(2)}-\dfrac{\laplacian{\varphi^{(2)}}}{a^2}\nonumber \\&+\qty[V_{\phi\phi}+2\dfrac{\dot{H}}{H}\qty(3H-\dfrac{\dot{H}}{H}+2\dfrac{\ddot{\phi}}{\dot{\phi}})]\varphi^{(2)}=D \,,
\end{align}
where
\begin{align}
    D&= g (\textbf{E} \cdot \textbf{B} )+ R + \dot{\phi}\qty(\dot{s}+\dot{s}_{F}) - 2 V_{\phi}\qty(s + s_{F}) \nonumber \\&+ \dfrac{\dot{\phi}}{2H}\qty[\qty(Q+Q_{F})-16 \pi G V (s + s_F)] \,,
   \end{align} 
  and
    \begin{align}
    R&=-\dfrac{1}{2}V_{\phi\phi\phi}\varphi^2-2\dot{\phi}\alpha\dot{\alpha}+\dot{\alpha}\dot{\varphi}+\dfrac{2}{a^2}\alpha\laplacian{\varphi}-2V_{\phi\phi}\alpha\varphi\nonumber \\&+\dfrac{\dot{\phi}}{2a}\alpha_{,\hspace{0.05cm}i}\beta^{,\hspace{0.05cm}i}-\dfrac{1}{4}V_{\phi}\abs{\pmb{\nabla}\beta}^2+\dfrac{1}{a^2}\alpha_{,\hspace{0.05cm}i}\varphi^{,\hspace{0.05cm}i}-\dfrac{H}{a}\beta_{,\hspace{0.05cm}i}\varphi^{,\hspace{0.05cm}i}\nonumber \\&-\dfrac{1}{2a}\dot{\varphi}\laplacian{\beta}+\dfrac{\dot{\phi}}{4}\beta_{,\hspace{0.05cm}i}\dot{\beta}^{,\hspace{0.05cm}i}-\dfrac{1}{2a}\varphi_{,\hspace{0.05cm}i}\dot{\beta}^{,\hspace{0.05cm}i}-\dfrac{1}{a}\beta_{,\hspace{0.05cm}i}\dot{\varphi}^{,\hspace{0.05cm}i} \,.
\end{align}
\\
We have therefore obtained perturbative equations, for the pseudo-scalar inflaton fluctuations and metric fluctuations, to second order including the contribution given by the gauge fields.

\section{Comoving observers}
\label{Sec5}

To obtain the gauge field's quantum backreaction, we have now to choose the observer w.r.t. such backreaction has to be evaluated. While for late-time cosmological observables the choice of the right observer depends on how an observation is done, for the case of a single field inflationary model we have a well-defined, and physically motivated, choice. During such an early universe phase, the only dynamical matter field that we can employ as a clock is the inflation field itself, hence the quantum backreaction should be evaluated with respect to the observer that uses such a clock.
Namely, we have to consider the so-called comoving observer, i.e. the observer who sees a homogeneous inflaton field for which $\varphi={\varphi}^{(2)}=0$. 
To evaluate the backreaction as experienced by the above observer in our framework, let us first note that in \cite{Finelli:2011cw,Marozzi:2013uva} was shown how in a single field model of inflation, taking into consideration the scalar sector and with no coupling between the inflaton and other fields, the class of comoving observers do not see on the effective Hubble rate any quantum backreaction effect, in the LW limit, for any potential $V(\phi)$ and to all orders in the slow-roll parameters. 
Therefore, for studying the total backreaction in our particular model, it is enough to take in consideration only the backreaction coming from the gauge fields. 

In our framework, the gauge fields contribute only to the second-order equations of motion, while do not change the dynamic of first-order perturbations. As a consequence, hereafter we can neglect the first-order square scalar contributions to the backreaction and focus only on the second-order gauge fields contribution to
Eq. (\ref{5.1.1-2}).  

Moving from a general coordinate system to the barred one, we can determine the expressions associated with equation (\ref{eq:media}) using the following coordinate transformation up to
second order \cite{Bruni:1996im}
\begin{equation}\label{eq:coordtrasf}
    x^{\mu} \rightarrow \tilde{x}^{\mu}=x^{\mu}+\epsilon^{\mu}_{(1)}+\dfrac{1}{2}\qty(\epsilon^{\nu}_{(1)}\partial_{\nu}\epsilon^{\mu}_{(1)}+\epsilon^{(\mu)}_{(2)}) \,,
\end{equation}
where $\epsilon^{\mu}_{(1)}$ and $\epsilon^{\mu}_{(2)}$ are the first order and second order generators, given by
{
\begin{equation}\epsilon^{\mu}_{(1)}=\qty(\epsilon^0_{(1)},\partial^i\epsilon_{(1)}+\epsilon^i_{(1)})\, , \quad \epsilon^{\mu}_{(2)}=\qty(\epsilon^0_{(2)},\partial^i\epsilon_{(2)}+\epsilon^i_{(2)}) \,,
\end{equation}}
with $\partial_i \epsilon^i_{(1)}=\partial_i \epsilon^i_{(2)}=0$.
\\
They enter in the associated first-order and second-order gauge transformations of a 
general tensor $T$ as (see, for example \cite{Marozzi:2010qz} for the detailed transformation of inflaton and metric fluctuations)
\begin{subequations}
\begin{align}
T^{(1)} \rightarrow \tilde{T}^{(1)}&=  T^{(1)}-L_{\epsilon_{(1)}} T^{(0)} \,,
\label{GGT1}\\
T^{(2)} \rightarrow \tilde{T}^{(2)}&= T^{(2)}-L_{\epsilon_{(1)}} T^{(1)}+\non\\&+\tfrac{1}{2}\left( L^2_{\epsilon_{(1)}} T^{(0)}-L_{\epsilon_{(2)}} T^{(0)}\right)\,,
\label{GGT2}
\end{align}
\end{subequations}
where $L_{\ep_{(1,2)}}$ is the Lie derivative respect the vector 
$\epsilon^\mu_{(1,2)}$.

As said, in our framework we can neglect first-order scalar contribution in our calculations (they have no impact on the backreaction). By neglecting them, we then obtain the following simplified 
gauge transformation to second order for a general scalar $A$ 
\begin{equation}
A^{(2)} \rightarrow \bar{A}^{(2)}=A^{(2)}-
\frac{1}{2} \epsilon^{0}_{(2)}\dot{A}^{(0)}\,,
\end{equation}
while for the second-order metric scalar perturbations  
we have
\begin{subequations}
\begin{eqnarray}
\bar{\alpha}^{(2)} &=& \alpha^{(2)} - \frac{1}{2} \dot \epsilon^0_{(2)}\,,
\\
\bar{\beta}^{(2)} &=& \beta^{(2)} - \frac{1}{a} \epsilon^0_{(2)} + a \dot \epsilon_{(2)}\,,
\\
\bar{\psi}^{(2)} &=&\psi^{(2)}+\frac{H}{2}\epsilon_{(2)}^0+\frac{1}{6}\nabla^2 \epsilon_{(2)}
\,,
\label{GTpsi}
\\
\bar{E}^{(2)} &=&E^{(2)}-\epsilon_{(2)}\,.
\end{eqnarray}
\end{subequations}
The observer which sees a homogeneous inflaton field must have $\bar{\varphi}^{(2)}=0$, hence the $\bar{x}^\mu$ reference system will correspond to the UFG.
The UFG conditions starting from the UCG are then obtained by choosing 
\begin{equation}
\epsilon^0_{(2)}=2 \frac{\varphi^{(2)}}{\dot{\phi}}
\quad,\quad
\ep_{(2)}=\int {\rm dt} \left(\frac{1}{a^2} \epsilon^0_{(2)}-\frac{1}{a}\beta^{(2)}\right) \,.
\label{UFG-UCG}
\end{equation}

By taking into account only the purely second-order terms and working in the long wavelength limit, we then have (by substituting Eq. (\ref{UFG-UCG}) in Eq. (\ref{GTpsi})) 
\begin{equation}
    \bar{\psi}^{(2)}=\dfrac{H}{\dot{\phi}}\varphi^{(2)} \,.
\end{equation}
The contribution that goes in the backreaction of Eq.~\eqref{5.1.1-2} is then given by
\begin{equation}\label{eq:psicomov}
    -\dfrac{2}{H}\langle \dot{\bar{\psi}}^{(2)} \rangle=-\dfrac{2}{H}\biggl< \qty[\qty(\dfrac{\dot{H}}{\dot{\phi}}-H\dfrac{\ddot{\phi}}{\dot{\phi}^2})\varphi^{(2)}+\dfrac{H}{\dot{\phi}}\dot{\varphi}^{(2)}] \biggr> \,.
\end{equation}

\section{Second order solution in $U(1)$-axion inflation}
\label{Sec6}

To evaluate the gauge field backreaction we have now to solve the equation of motion for the pseudo-scalar inflaton to second order within the UCG. 

For our purpose, it is enough to take
in consideration only the gauge fields contribution to $\varphi^{(2)}$. Therefore, starting from Eq.(\ref{eq:5.2.1-1}), working at leading order in the slow-roll parameters and in the long-wavelength approximation, we obtain
\begin{align}
   &\ddot{\varphi}^{(2)} + 3 H \dot{\varphi}^{(2)}  +\left[3 \dot{H}-\frac{3}{2} H \frac{\ddot{H}}{\dot{H}}\right] \varphi^{(2)} 
   \simeq  g (\textbf{E} \cdot \textbf{B} ) + \dot{\phi} \dot{s}_F
   \nonumber \\& -2V_{\phi} s_F
   +\dfrac{\dot{\phi}}{2H}\qty(Q_F-\frac{2}{M_{Pl}^2} V  s_F) \,.
\end{align}
\\
The above equation can also be written as
\begin{align}
   &\ddot{\varphi}^{(2)} + 3 \dv{t}\qty(H \varphi^{(2)})
   -\frac{3}{2} H \frac{\ddot{H}}{\dot{H}} \varphi^{(2)} \simeq g (\textbf{E} \cdot \textbf{B} )+
    \dot{\phi} \dot{s}_F \nonumber \\&
    -2V_{\phi}\,s_F
    +\dfrac{\dot{\phi}}{2H}\qty(Q_F-\frac{2}{M_{Pl}^2} V  s_F) 
    \label{GF2ordContrphi2}
    \,.
\end{align}
\\
At this point, to obtain the gauge fields contribution to $\langle \varphi^{(2)} \rangle$, we need to do first the v.e.v. of Eq.(\ref{GF2ordContrphi2}) and then integrate the obtained equation with respect to time. 
The first step produces the following dynamical equation at leading order in the slow-roll parameters
\begin{align}
    &\dv{t}\qty(H \langle \varphi^{(2)}\rangle )-\frac{H}{2} \frac{\ddot{H}}{\dot{H}} \langle \varphi^{(2)}
    \rangle 
    \nonumber\\& \simeq\frac{g}{3} \langle \textbf{E} \cdot \textbf{B} \rangle
    - \frac{\dot{\phi}}{6 H M_{Pl}^2}  \dfrac{\langle \textbf{E}^2+\textbf{B}^2 \rangle}{2}
    \label{vevGF2ordContrphi2}
    \,,
\end{align}
where we assumed $\langle \ddot{\varphi}^{(2)} \rangle \sim (\dot{H}/H) \langle \dot{\varphi}^{(2)} \rangle$, as it can be verified once we obtained the solution for $\langle \varphi^{(2)} \rangle$.
Furthermore, we used that
\begin{equation}
    \langle Q_F \rangle = -8 \pi G \dfrac{\langle \textbf{E}^2+\textbf{B}^2 \rangle}{2} \,,
\end{equation}
and the fact that 
the vacuum expectation value of $s_F$ vanishes because $\langle \pmb{\nabla}\cdot (\textbf{E} \times \textbf{B}) \rangle =0$ for the isotropy of the background. Namely, we have 
\begin{equation}
    \langle s_F \rangle=-\Big \langle \dfrac{1}{\laplacian}[\pmb{\nabla}\cdot (\textbf{E} \times \textbf{B})] \Big \rangle =0 \,.
\end{equation}

In order to solve Eq.\eqref{vevGF2ordContrphi2}, let us now do the following ansatz
\begin{equation}
    \langle \varphi^{(2)}\rangle = \frac{\g(H)}{\dot{H}} 
    \frac{\dot{\phi}}{6 H M_{Pl}^2} \Ene+\frac{\s(H)}{\dot{H}}
    {g\over 3} \Hel\,,
\end{equation}
where $\g(H)$ and $\s(H)$ are unknown functions to be determined.
At leading order in slow-roll, and in the chaotic $m^2 \phi^2$  inflation limit~\footnote{For this case we have $\frac{\ddot{H}}{\dot{H} H}={\cal O}\left(\frac{\dot{H}}{H^2}\right)^2$. We use this limit because, as described in Sec.~\ref{sec7},  we will evaluate the backreaction considering natural inflation with $f\gg M_{Pl}$. In this case, natural inflation is really well approximated by this chaotic potential.}, we obtain then the following differential equations for such functions: 
\begin{subequations}
\begin{align}
     &H\pa_H\g +\g H \pa_H \ln \r_{\tx{EM}}=-1\,,\\
     \non\\
     &H\pa_H\s +\s(1  + H \pa_H \ln \r_{\tx{Hel}})=1\,,
\end{align}
\end{subequations}
with solution
\begin{subequations}
\begin{align}
   \g(H)&=- {1\over \r_{\tx{EM}}} \int {\r_{\tx{EM}} \over H}  \dd H\,, \\
   \s(H)&={1\over H \r_{\tx{Hel}}} \int \r_{\tx{Hel}}  \dd H \,,
\end{align}
\end{subequations}
where we have defined 
\begin{eqsplit}
&\r_{\tx{EM}} \equiv \Ene \,, \qquad \r_{\tx{Hel}} \equiv \Hel \,. \non
\end{eqsplit}
Finally, we obtain 
\beq
\VEV{\vp^{(2)}}= -\frac{1}{6} \frac{\dot{\phi}}{H \dot{H} M_{Pl}^2} \int {\r_{\tx{EM}} \over H}  \dd H + {g \over 3} \frac{1}{H \dot{H}} \int \r_{\tx{Hel}}  \dd H
\,. \label{phi2dotFin}
\eeq

At this point, we are able to evaluate the effective Hubble parameter
\begin{align}
H^2_{\eff}=&\le({1\over a_{\eff}} \pdv{a_{\eff}}{A_0} \ri)^2 \non\\ =&\le[1-{2 \over \fd} \le({\Hd \over H}-{\ddot{H}\over 2\Hd}\ri) \VEV{\vp^{(2)}}- {2 \over \fd}\VEV{\dot{\vp}^{(2)} } \ri] \non\\
\simeq & \, H^2 \le[1-{2 \over \fd} \le({\Hd \over H} \VEV{\vp^{(2)}}+ \VEV{\dot{\vp}^{(2)} } \ri) \ri]\,.
\end{align}
The contribution $\left< \dot{\varphi}^{(2)} \right>$ can be explicitly calculated. Starting from Eq. (\ref{phi2dotFin}), and at leading order in slow-roll,
we obtain
\begin{equation}
  \left< \dot{\varphi}^{(2)} \right> \simeq -\frac{\dot{H}}{H} \left<{\varphi}^{(2)} \right>  -\frac{1}{6} \frac{\dot{\phi}}{H^2 M_{Pl}^2}  \r_{\tx{EM}}  + {g \over 3} \frac{1}{H}  \r_{\tx{Hel}}  \,.
\end{equation}
This last equation, together with Eq. (\ref{phi2dotFin}), shows the validity of the previous assumption $\langle \ddot{\varphi}^{(2)} \rangle \sim (\dot{H}/H) \langle \dot{\varphi}^{(2)} \rangle$.

Finally, we have 
\beq
H^{2}_\eff= H^{2}\le[1+ {1\over 3 H^2} \le( \frac{1}{M_{Pl}^2} \Ene -{g^2 \over \xi } \Hel  \ri) \ri] \,,
\eeq
or, considering that we are in the weak-backreaction regime, we can also write 
\beq
H_\eff= H\le[1+ {1\over 6 H^2}\le(\frac{1}{M_{Pl}^2} \Ene -{g^2 \over \xi } \Hel  \ri) \ri] \,,
\label{H_eff}
\eeq
where we define $\xi\equiv\frac{g\dot{\phi}}{2H}$.


\section{Numerical analysis}
\label{sec7}

Let us now investigate the magnitude of the obtained 
gauge-invariant backreaction effect, which includes both metric and matter scalar perturbations. To this aim, we perform the integration of the homogeneous field equation
\beq
 \ddot{\phi}+3H\dot{\phi}+V_{\phi}=0\,,
\eeq
written in terms of the effective number of e-fold $N_\eff$ defined by
\beq
\dd N_\eff \equiv H_\eff~ \dd t\,,
\label{Eff e-f}
\eeq
where $H_\eff$ is given by Eq.~\eqref{H_eff}.
The complete system of equations, in terms of $\xi$, reads then
\begin{subequations}
\begin{align}
&\xi'+\le(\tfrac{H'}{H}+3\tfrac{H}{H_\eff} \ri)\xi +\tfrac{g}{2}\tfrac{V_{,\f}}{H H_\eff}=0\,, \label{sys93a}\\
&\f'= \tfrac{2}{g}\tfrac{H}{H_\eff}\xi \label{sys93b}\,,
\end{align}
\end{subequations}
where now a prime states a derivative w.r.t. $N_\eff$.
The above integration can then be compared with what is usually done in literature \cite{Anber:2009ua}, i.e. the integration of the equation 
\beq
 \ddot{\phi}+3H\dot{\phi}+V_{\phi}=g \langle \textbf{E} \cdot \textbf{B} \rangle \,.\label{phi2BRV}
\eeq
The main difference between the two approaches is that in our case the backreaction effect is implicitly encoded in the e-fold number of Eq.\eqref{Eff e-f}, where we are taking into consideration also the induced scalar fluctuations.

The missing one ingredient to close the system is the
inflationary model, which we choose to be natural
inflation, first proposed in~\cite{Freese:1990rb,Adams:1992bn}. For this model we have
\begin{equation}
\label{pot:natinf}
   V(\phi) = \Lambda^4 \left[ 1 -\cos\left(\frac{\phi}{f}\right)\right] \,.
\end{equation}
 The correspondent slow-roll parameters are given by
\begin{align}
    \epsilon&=\dfrac{1}{16\pi G}\qty(\dfrac{V'}{V})^2=\frac{1}{2}\left(\frac{M_{Pl}}{f}\right)^2
\left[
    \frac{\sin(\phi/f)}{1-\cos(\phi/f)}\right]^2 \,,\\ 
    \eta&=\dfrac{1}{8\pi G}\qty(\dfrac{V''}{V})=
    \left(\frac{M_{Pl}}{f}\right)^2
    \frac{\cos(\phi/f)}{1-\cos(\phi/f)} \,,
\end{align}
and we then have the following relation 
\begin{equation}
    \eta=\epsilon-\frac{1}{2}\left(\frac{M_{Pl}}{f}\right)^2 \,.
\end{equation}

Assuming to be in the slow-roll regime $\epsilon\ll 1$, we also have $\eta\ll 1$ only if $f^2\gg M_{Pl}^2$. 
In the following, we indeed assume $f^2\gg M_{Pl}^2$ to have a model consistent with observation \cite{Planck:2018jri}. As mentioned, in this limit natural inflation can be well approximated by chaotic inflation.
The detailed study of the backreaction, when the condition $f^2\gg M_{Pl}^2$ does not hold, is postponed to future investigation.

The initial conditions necessary to solve the system can be calculated as done in the absence of backreaction, being the backreaction negligible at the initial instants of the inflationary phase.
To better compare our results with previous findings (e.g. \cite{Domcke:2016bkh}), we choose these initial conditions to obtain a 60 e-folds long inflationary phase including the backreaction as evaluated by Eq.(\ref{phi2BRV}). 
Let us then use the number of e-folds to find the value of the scalar field at the beginning of inflation. In the absence of backreaction, we have
\beq
N=\int_{t_{i}}^{t_f}H \dd t=\int_{\phi_{f}}^{\phi_i}\frac{V}{V_{\phi}}\dd \phi\, .
\eeq
To obtain $\phi_i$ we need the field value at the end of inflation, which is obtained with the condition $\epsilon=1$. Following \cite{Domcke:2016bkh}, we then obtain $60$ e-folds including backreaction, if without backreaction we require a $N=42$ e-folds long inflationary phase. Thus, we computed $\phi_i$ using $N=42$ and then applied it to find $\xi_i=-\frac{g}{2}\frac{V_{\phi}}{V}|_{\phi_i}~$\footnote{The equation of motion in the slow-roll regime without backreaction reads as
\beq
\phi'+ \frac{V_{\phi}}{V} = 0 \,.
\eeq
}. In addition, we considered the values of $g=60/M_{Pl}$ and 
$f=5 M_{Pl}$ throughout the numerical analysis.

\begin{figure*}[t]
\begin{minipage}{1\textwidth}
\subfigure[]{
\hspace{-15mm}
\includegraphics[scale=0.31]{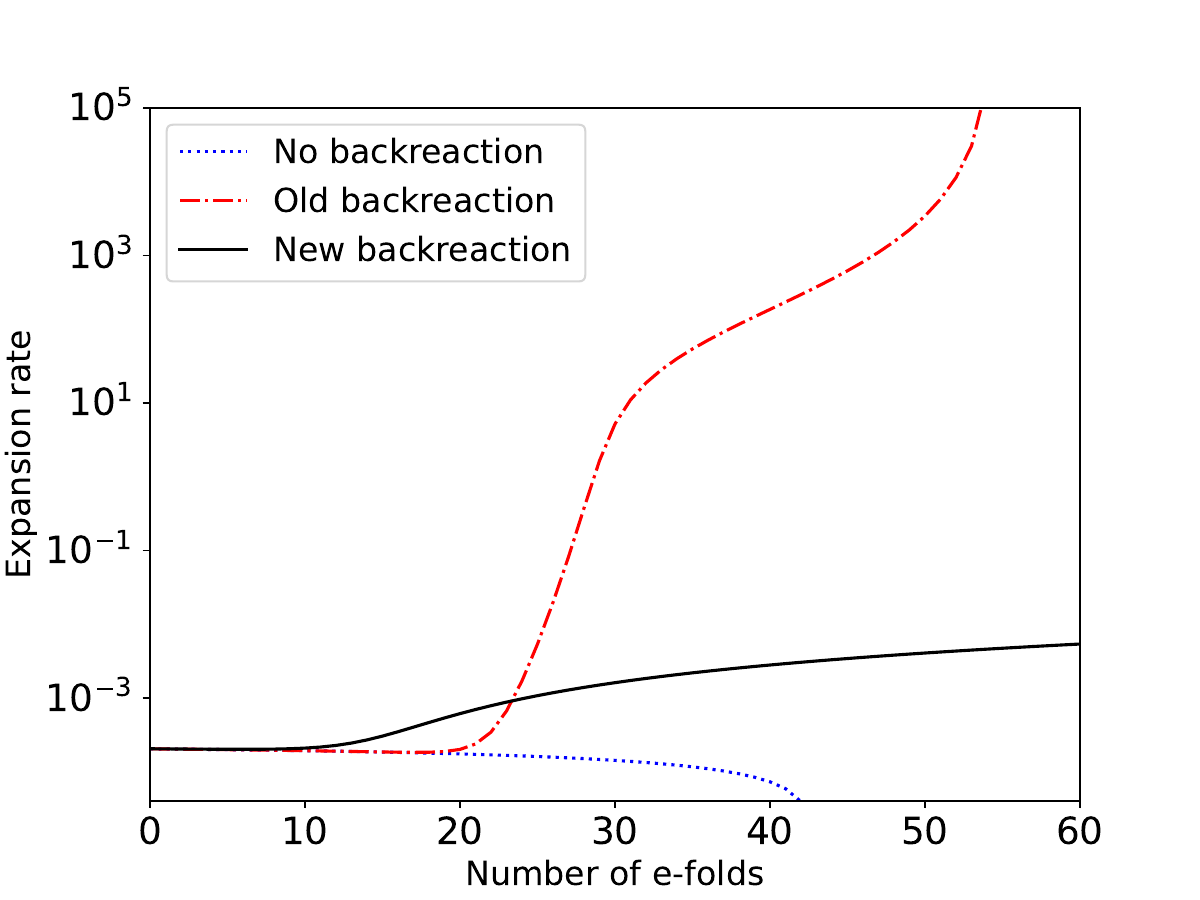}
\label{Fig1a}}
\subfigure[]{
\hspace{-9mm}
\includegraphics[scale=0.295]{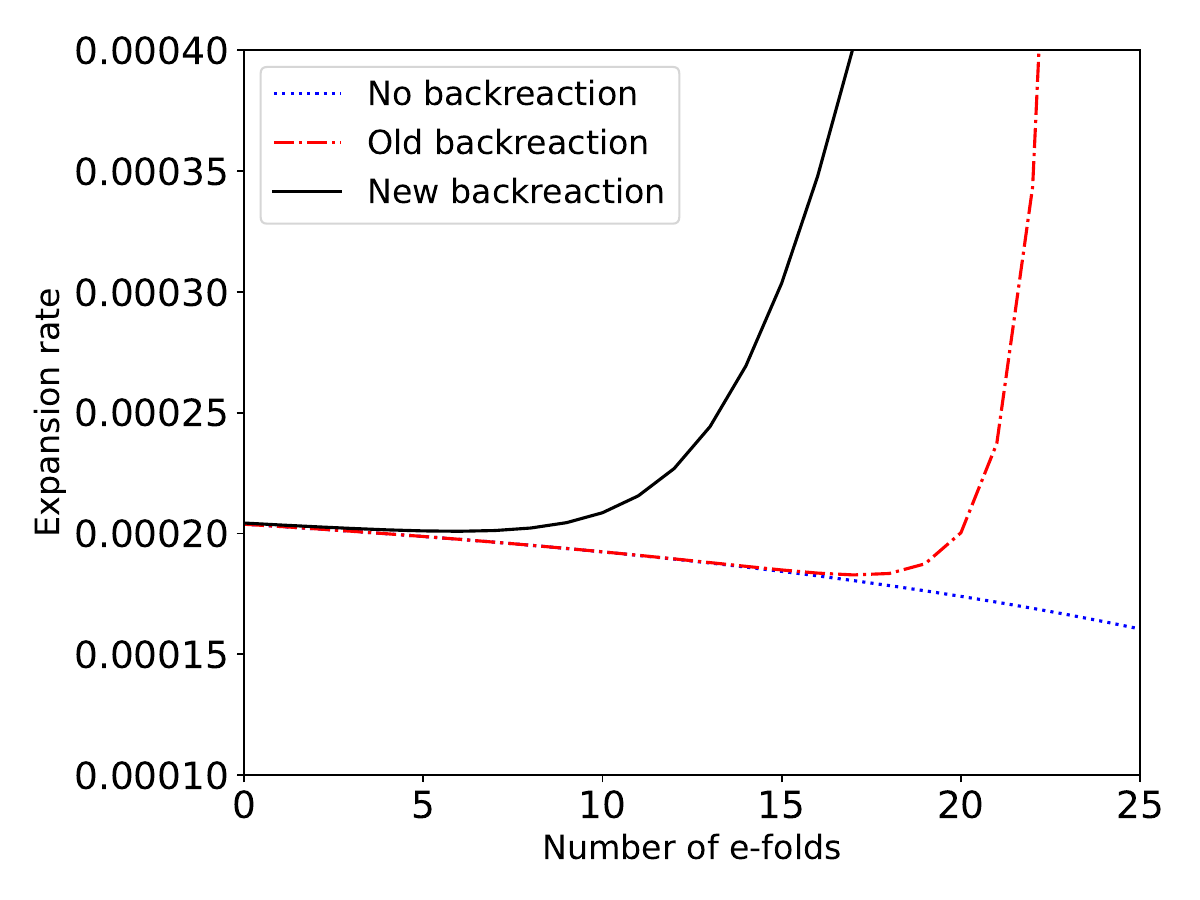}
\label{Fig1b}}
\hspace{-5mm}
\subfigure[]{
\includegraphics[scale=0.31]{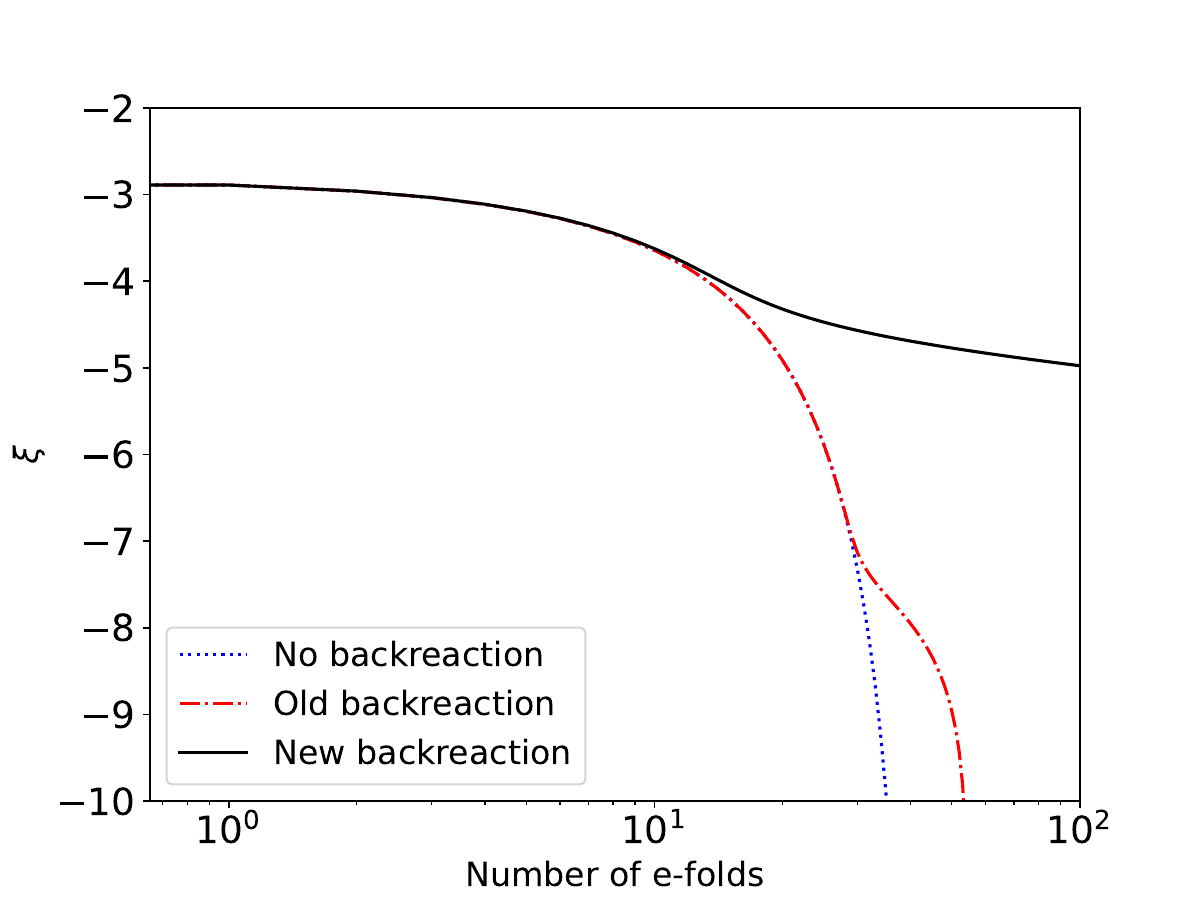}
\label{Fig1c}}
 \end{minipage}
 \caption{Time evolution for the expansion rate \textbf{(a, b)} and $\xi$ parameter \textbf{(c)}, considering $g=60/M_{Pl}$ and $f=5 M_{Pl}$. We show the results of solving the Friedmann and Klein-Gordon equations in the absence of backreaction (blue dotted lines), the case when only the gauge fields production is taken into account (red dot-dashed lines), and the current work in solid black lines, where we consider also metric and field scalar perturbations to calculate the effective expansion rate.}
\label{Fig:Hcsi}
\end{figure*}

Finally, one can integrate the system of Eqs.~\eqref{sys93a} and \eqref{sys93b} to obtain the solutions for $\phi$ and $\xi$, and then evaluate the consequent effective expansion rate as given by Eq.~\eqref{H_eff}.
The results are shown in black lines in \fig{Fig1a} and \fig{Fig1b}. For completeness, we also included in the figures the results of solving the Friedmann and Klein-Gordon equations, both in the absence of backreaction (blue dotted line) and in the case when only the gauge field production is taken into account for the backreaction
(red dot-dashed line), which means using \eq{phi2BRV} (see \cite{Domcke:2016bkh,Ballardini:2019rqh}).
Let us note that for our case, the impact of the backreaction 
on the effective expansion rate (\fig{Fig1a} and \fig{Fig1b}) becomes large before
 with respect to the case in which only the gauge field production is taken into account. 
 This is also noticed on the parameter $\xi$ (\fig{Fig1c}), which controls the influence of the gauge fields on the inflaton dynamics. It starts to change at $N\sim 15$ but evolves much slower compared with the standard dynamics and the other backreaction considered (this latter starts to differ from the standard dynamics at $N\sim 35$).

Let us now further discuss the behavior of \( H_\eff \) as a function of the effective number of e-folds. In both scenarios where we consider the influence of backreaction, we observe that \( H_\eff \) begins to increase with the effective number of e-folds after a certain point, whereas, as expected, in the background case it always decreases with the number of e-folds.
This behavior is a consequence of the rise of the backreaction,
which becomes significant enough (non-perturbative) to reverse the sign of the variation of \( H_\eff \) with respect to the number of e-folds. 
As well illustrated in \fig{Fig1b}, in our case this reversal occurs around \( N \sim 8 \).
This result implies that our approach may lose validity 
already at this stage,
and the evolution of \( H_\eff \) should be treated with caution even before it exhibits substantial deviations from its background value around \( N \sim 15-20 \) (see \fig{Fig1a} and \fig{Fig1b}).

\begin{figure*}[]
\begin{minipage}{1\textwidth}
\subfigure[]{\hspace{-10mm}
\includegraphics[scale=0.31]{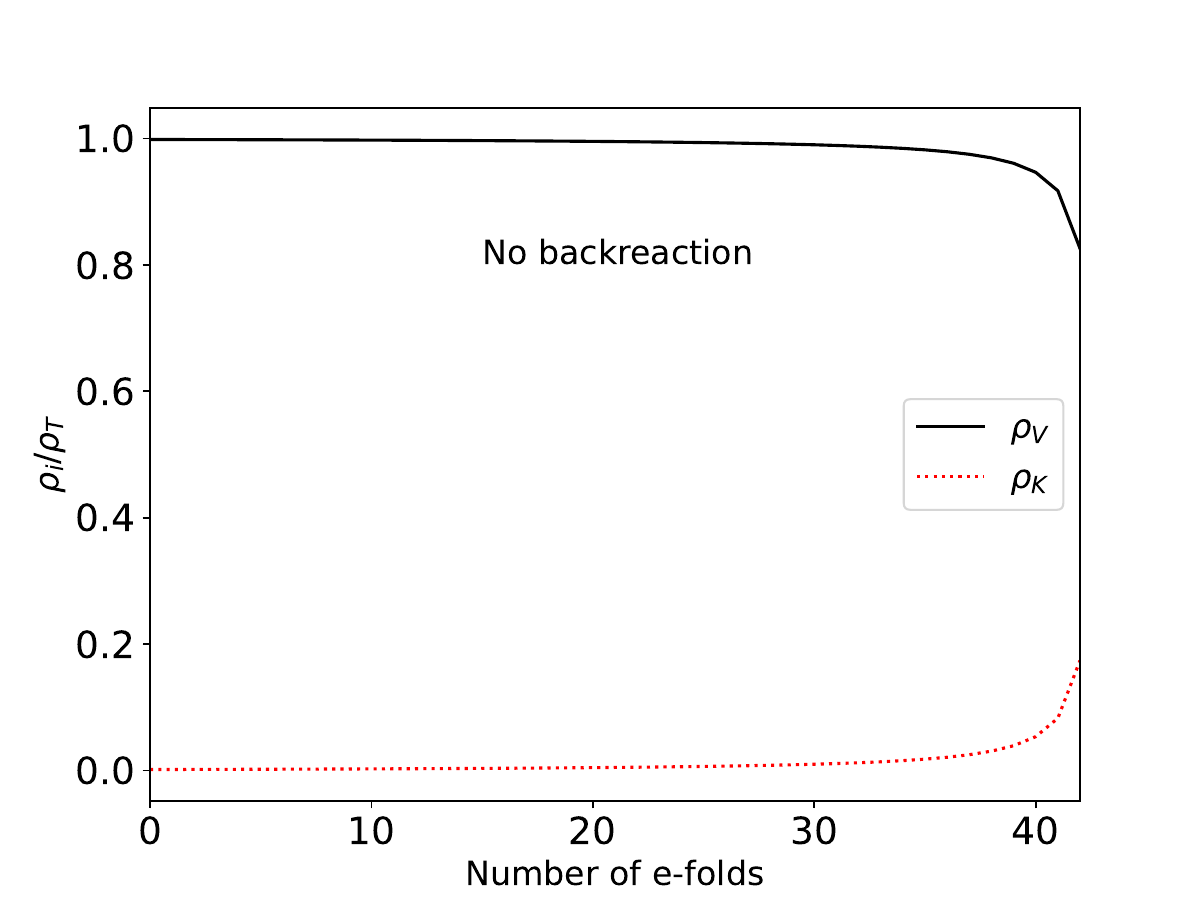}
\label{Fig:densitiesa}\hspace{-9mm}}
\subfigure[]{
\includegraphics[scale=0.31]{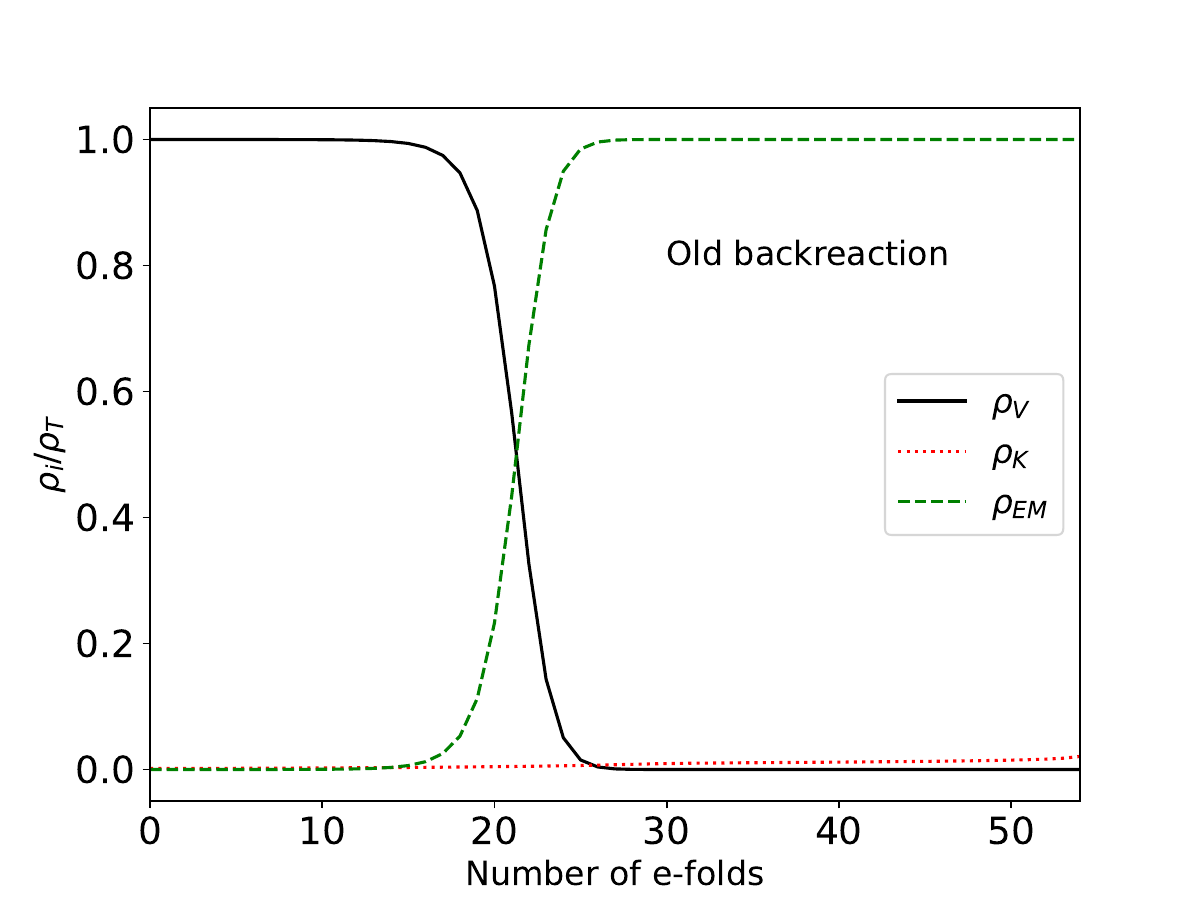}
\label{Fig:densitiesb}
\hspace{-9mm}}
\subfigure[]{
\includegraphics[scale=0.31]{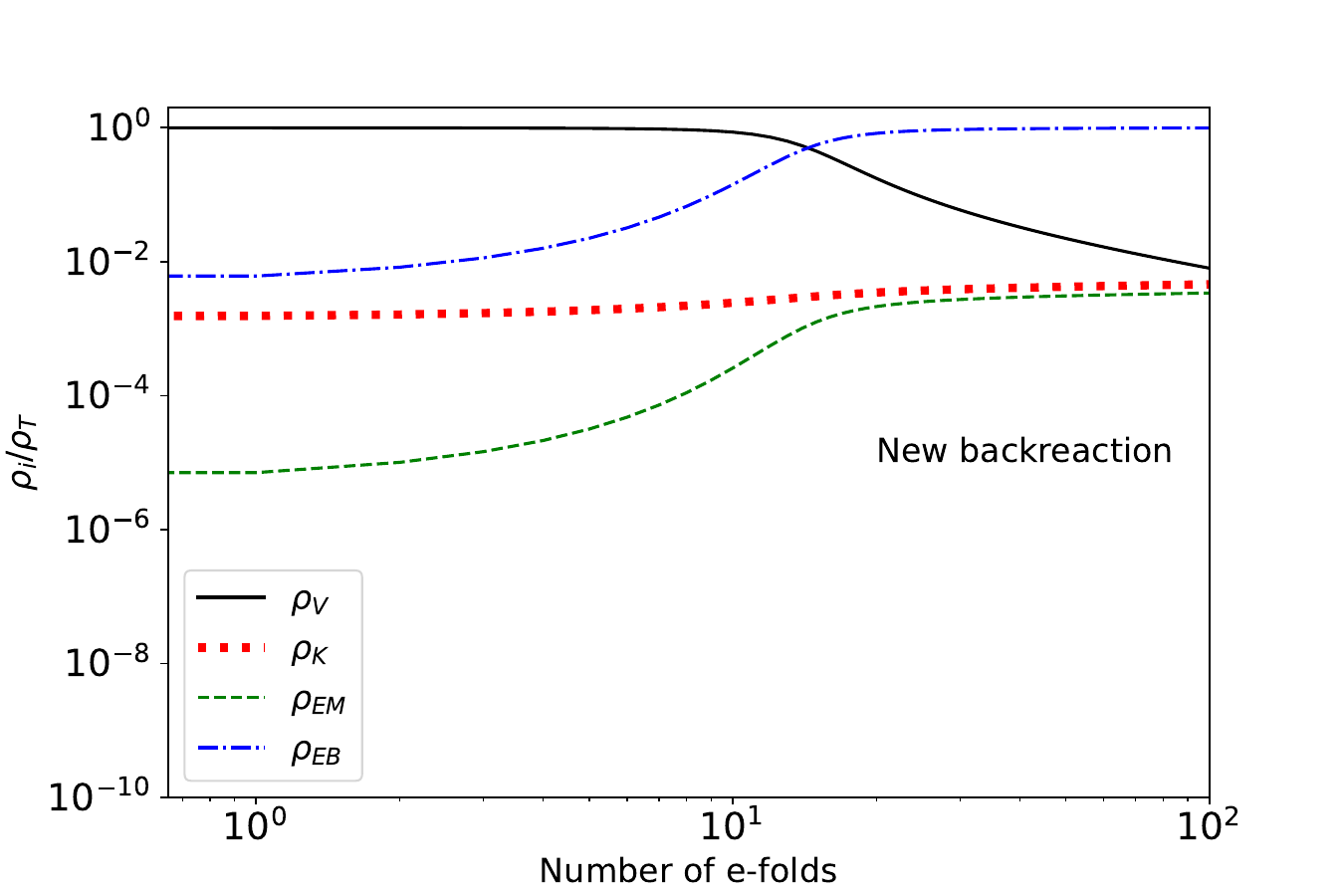}
\label{Fig:densitiesc}}
\end{minipage}
\caption{Relative density evolution for the cases: \textbf{(a)} no backreaction, \textbf{(b)} backreaction due to only gauge fields production, and \textbf{(c)} new effective backreaction which includes both field and metric scalar perturbations. In all cases, we considered $\rho_{\tx{V}}=V(\phi)$, $\displaystyle \rho_{\tx{k}}=\tfrac{\dot{\phi}^2}{2}$, $\displaystyle \rho_{\tx{EM}}=\tfrac{\VEV{{\bf E}^2+{\bf B}^2}}{2}$, and $\displaystyle \rho_{\tx{EB}}=-\tfrac{g^2 M_{Pl}^2}{\xi}\Hel$, where $\phi$ is the one obtained solving the corresponding system. When required, we considered $g=60/M_{Pl}$ and $f=5 M_{Pl}$.}
\label{Fig:densities}
\end{figure*}

Other interesting quantities to examine are the relative densities since we can identify when each becomes important during cosmic evolution. Rewriting \eq{H_eff} as
\beq
3 M_{Pl}^2 H^{2}_\eff= \frac{\dot{\phi}^2}{2}+ V(\phi)+\Ene -{g^2 
 M_{Pl}^2 \over \xi } \Hel  \,,
\label{ResFig2NewApp1}
\eeq
we can identify
\begin{align}
\rho_{\tx{V}}&\equiv V(\phi), \qquad \quad \rho_{\tx{k}}\equiv \frac{\dot{\phi}^2}{2}, \qquad \quad \rho_{\tx{EM}}\equiv \Ene,\nonumber \\
\rho_{\tx{EB}}&\equiv -{g^2 M_{Pl}^2 \over \xi } \Hel, \qquad \mathrm{and} \quad \,\,\,  ~\quad \rho_T=3 M_{Pl}^2 H^{2}_\eff\,. 
\label{ResFig2NewApp2}
\end{align}
The evolution of the relative densities with the number of e-folds can be seen in \fig{Fig:densities}. We show three cases: in \fig{Fig:densitiesa} one can see their evolution in the absence of backreaction. Indeed, we have the classic slow-roll approximation for the inflaton dynamics, with the potential energy dominating the kinetic term until the very end of inflation. In the second case, shown in \fig{Fig:densitiesb}, the gauge field production is the only source for the backreaction. Here, one can observe the inflaton energy's decay into the gauge field production by increasing the $\rho_{\tx{EM}}$ component. 
 Lastly, in \fig{Fig:densitiesc}, we show the new results associated with our approach as described in Eqs.(\ref{ResFig2NewApp1})
and (\ref{ResFig2NewApp2}).
We always start with the potential energy dominating the other densities. However, it is crucial to note that the contribution of the helicity term, $\rho_{\tx{EB}}$, is consistently larger than $\rho_{\tx{EM}}$ and $\rho_{\tx{k}}$ from the beginning. Notably, $\rho_{\tx{EB}}$ exceeds the potential energy at an early stage ($N\sim 15$). 
Furthermore, from $N\sim 15$ we also have that $\rho_{\tx{EM}}$ becomes comparable to the kinetic term. 

\section{Discussion and Conclusions}
\label{Conc}

In this manuscript, we have reconsidered the cosmological backreaction problem for a pseudo-scalar field $\phi$ driving inflation coupled to gauge fields via a Chern-Simons-like operator, the so-called $U(1)$-axion inflation model (see \cite{Anber:2009ua, Sorbo:2011rz, Anber:2012du, Barnaby:2011vw}).
We have built our investigation starting from two pillars.
First, from the idea that to obtain a meaningful description of a physical phenomenon, like backreaction, in a cosmological context one should always consider gauge invariant observables. Otherwise, the effect of perturbations might be attributed to a gauge artifact, making any physical interpretation extremely challenging. 
Second, since cosmological backreaction arises from the non-linearity of Einstein's equations, for a consistent description within perturbation theory, one has to include both matter and metric fluctuations in the framework, being all these fluctuations coupled among them. 
Therefore, starting from the covariant and gauge invariant approach 
introduced in \cite{Gasperini:2009wp, Gasperini:2009mu}, we have studied the effective evolution of the scalar factor as measured by an observer comoving with the inflaton field. The analytical expression obtained, valid at the leading order in the slow-roll approximation, is shown in Eq.\eqref{H_eff}, while the backreaction impact on the model and space-time evolution is quantified in \fig{Fig:Hcsi} and \fig{Fig:densities}.

The backreaction obtained within our gauge-invariant approach, considering inflaton and scalar metric fluctuations, is quite different with respect to the one obtained with standard approaches, which consider only the gauge field contribution including, at maximum, the inflaton fluctuations.
Looking at \fig{Fig1a} and \fig{Fig1b} we can note how the backreaction begins to be significant well before the standard case. Such backreaction became non-perturbative after $N\sim 8-15$ e-folds, while in the standard case, one has a non-perturbative impact of the backreaction only at  $N\sim 18 - 23$ e-folds, depending if we are looking to $H_\eff$ or its derivative.

In \fig{Fig:densities} we show the evolution of the relative densities with the number of e-folds. 
An interesting feature is that differently from the old backreaction approach, now the main contribution to the energy density evolution is explicitly due to the helicity integral contained in $\rho_{EB}$.
Indeed, such a contribution becomes larger than the one associated with the potential energy for $N \sim 15$ and, as a consequence, the backreaction becomes dominant and non-perturbative, this confirms the above considerations based on \fig{Fig:Hcsi}.

Correspondingly, \fig{Fig:Hcsi} and \fig{Fig:densities} demonstrate that the cosmological backreaction became non-perturbative already after $N\sim 8-15$ e-folds.
When this happens our assumption of considering the gauge field contribution as a perturbative contribution is not valid anymore. Having said that, to go beyond this weak backreaction regime, considering all the possible contributions, one possibility could be to define an effective background for the gauge fields' contribution and write a perturbative theory on top of it, to take into consideration the coupling between the scalar fluctuations of the matter/gravity sector with the gauge fields effective fluctuations consistently.
The above approach is necessary to consider, in such a case, not only the backreaction of the gauge fields on the inflationary dynamics but also the backreaction that the scalar fluctuations of the matter/gravity sector have on the gauge fields' production. This non-trivial issue will be addressed in a forthcoming paper. 
Moreover, this above extension would be crucial to accurately evaluate the impact of backreaction also on the perturbation spectra.
In this regard, it is noteworthy that, as discussed in \cite{Durrer:2024ibi}, scalar metric perturbations significantly affect the spectrum of the comoving curvature perturbation.

To conclude, let us underline how the approach proposed in this manuscript takes into consideration consistently both the scalar fluctuations of the matter/metric sector and their coupling with the gauge fields.  This approach gives us the possibility to study the weak backreaction regime and understand when this is not valid anymore, without missing part of the effect coming from scalar-induced perturbations. 
 Finally, looking at \fig{Fig:densitiesc} we see how the contribution coming from metric perturbations, associated with the value of the gauge fields energy density $\rho_{EM}$, is negligible throughout the perturbative regime~\footnote{We thanks Marco Peloso for drawing our attention to this interesting point.}.
However, looking at the growing behavior of the energy density $\rho_{EM}$, one could conjecture that such a contribution could become non-negligible going towards the strong backreaction regime.
 This requires further investigations that we postpone to future work.\\

\section*{Acknowledgements}
 We wish to thank Chiara Animali, Matteo Braglia, Ruth Durrer, Marco Peloso, and Lorenzo Sorbo for useful comments and discussions.
 DCG, PC, GM and SSC are supported in part by the Istituto Nazionale di Fisica Nucleare (INFN) through the Commissione Scientifica Nazionale 4 (CSN4), GM and SSC under the Iniziativa Specifica (IS) Theoretical Astroparticle Physics (TAsP) and the IS Quantum Fields in Gravity, Cosmology and Black Holes (FLAG), while DCG and PC under the IS TAsP.
The work of GM was partially supported by the research grant number 2022E2J4RK “PANTHEON: Perspectives in Astroparticle and Neutrino THEory with Old and New messengers” under the program PRIN 2022 funded by the Italian Ministero dell’Università e della Ricerca (MUR). 
SSC acknowledges support from the Fondazione Cassa di Risparmio di Trento e Rovereto (CARITRO Foundation) through a Caritro Fellowship (project ``Inflation and dark sector physics in light of next-generation cosmological surveys''). The work of PC is supported by a Della Riccia foundation grant. PC is thankful to the CERN Theoretical Physics Department for the hospitality and financial support provided during the development of
this project. PC acknowledge support by the MIUR Progetti di Ricerca di Rilevante Interesse Nazionale (PRIN) Bando 2022 - grant 20228RMX4A, funded by the European Union - Next generation EU, Mission 4, Component 1, CUP C53D23000940006.
\newpage
\appendix
\input{Appendix}

\bibliographystyle{apsrev4-1}

\bibliography{bibliography}

\end{document}

%% file: Appendix.tex
\begin{appendices}
\onecolumngrid

\section{- $\,$ Gauge fields contribution}
\label{app.A}
The Fourier mode functions $A_\pm$ of the gauge fields satisfy the following equation of motion
\beq\label{eom_A}
\frac{{\rm d}^2 }{{\rm d} \tau^2} A_\pm(\tau,k) + \left(k^2 \mp k g \f^\prime \right) A_\pm(\tau,k)=0\,.
\eeq
Under the assumption of de Sitter expansion for the background (i.e. $a(\t)=-1/(H \t)$ with $\t<0$, $H= \text{const.}$ and $\dot{\f}= \text{const.}$), we can rewrite the above equation in terms of the constant parameter $\xi \equiv g  \f^\prime /(2 a(\t)  H)= g  \dot{\f} / ( 2 H) $. In this approximation the analytical solution of Eq.(\ref{eom_A}) is given in terms of \textit{Whittaker}  $W-$ functions
\beq\label{mode_funct_A}
A_\pm(\tau,k)= \frac{1}{\sqrt{2k}} e^{\pm \pi \xi/2} W_{\pm i \xi , \frac{1}{2}} (-2 i k \tau)\,.
\eeq

The gauge fields contribution can be given in terms of two physical quantities. The \textit{vacuum expectation value} of the energy density of the gauge fields
\begin{align}\label{eq:4.1-4}
   \dfrac{\langle \textbf{E}^2+\textbf{B}^2 \rangle}{2}=\bigintsss &\dfrac{dk}{(2\pi)^2a^4}k^2 \Bigl[\abs{A'_+}^2+\abs{A'_-}^2+k^2 \bigl(\abs{A_+}^2+\abs{A_-}^2\bigr)\Bigr] \,,
\end{align}
and the so-called helicity integral
\begin{equation}\label{eq:4.1-5}
    \langle \textbf{E} \cdot \textbf{B} \rangle=-\bigintsss \dfrac{dk}{(2\pi)^2a^4}k^3 \pdv{}{\tau}\qty(\abs{A_+}^2-\abs{A_-}^2) \,.
\end{equation}
These integrals diverge for large $k$, as usual in quantum field theories in curved backgrounds \cite{Birrell:1982ix, Parker:2009uva}.

One can perform a proper renormalization procedure by subtracting to the bare results of the energy density and of the helicity integral their adiabatic counterparts as described in \cite{Animali:2022lig}.

The final renormalized result for the energy density is \cite{Animali:2022lig}
\begingroup
\setlength{\abovedisplayskip}{20pt}
\beq
\label{energybetascheme}
\begin{split}
\Ene=&\, \frac{2 H^4}{960 \pi^2} +\frac{H^4 \xi^2\left(-1185 \xi^4 + (330+4\sqrt{15})\xi^2+435+4\sqrt{15}\right) }{960 \pi ^2 \left(1+\xi ^2\right)}\\
   &-\frac{3 H^4 \xi ^2   \left(5 \xi ^2-1\right) \log \left(15/4\right)}{64 \pi ^2}+\frac{ H^4\xi    \left(30 \xi
   ^2-11\right) \sinh (2 \pi  \xi )}{64 \pi ^3}\\
   &-\frac{3  H^4  \xi^2\left(5 \xi ^2-1\right) (\psi(-1-i \xi )+\psi(-1+i \xi ))}{32 \pi ^2}\\
   &+\frac{3 i  H^4 \xi^2
   \left(5 \xi ^2-1\right) (\psi ^{(1)}(1-i \xi )-\psi ^{(1)}(1+i \xi )) \sinh (2 \pi  \xi )}{64 \pi ^3}\,,
   \end{split}
   \eeq
\endgroup
while for the helicity integral we obtain
\beq\label{helicitybetascheme}
\begin{split}
\br\ve{E} \cdot \ve{B}\ke= &\,\frac{H^4 \xi  \left(705 \xi^2-330 -\sqrt{15}\right)}{240 \pi ^2}+\frac{3 H^4 \xi  \left(5 \xi ^2-1\right) \log \left(15/4\right)}{32 \pi ^2}\\
&+\frac{3 H^4 \xi  \left(5 \xi ^2-1\right) (\psi(1-i \xi )+\psi(1+i \xi ))}{16 \pi ^2}\\
&+\frac{3 i H^4 \xi  \left(5 \xi ^2-1\right) (-\psi ^{(1)}(1-i \xi )+\psi ^{(1)}(1+i \xi )) \sinh (2 \pi  \xi )}{32 \pi ^3}\\
&+\frac{H^4 \left(11-30 \xi ^2\right) \sinh (2 \pi  \xi )}{32 \pi ^3}\,.
\end{split}
\eeq
where $\psi(x)$ is the Digamma function and $\psi^{(1)}(x)\equiv \tx{d}\psi(x)/ \tx{d} x $\,.

\newpage
\section{- $\,$ Einstein and energy-momentum tensor to second order}
\label{app.B}

Within the perturbative framework introduced in the main text, the Einstein tensor expanded to second order in the UCG is given by (see \cite{Finelli:2003bp}):

\begin{eqnarray}
\label{E-tensor}
G^0_0 &=& 
G^{0 (0)}_{0} + \delta G^{0 (1)}_{0} + \delta
G^{0 (2)}_{0} \nonumber \\
&=&
- 3 H^2 - \frac{H}{a} \nabla^2 \beta + 6 H^2 \alpha
\nonumber \\
& & - \frac{H}{a} \nabla^2 \beta^{(2)} + 6 H^2 \alpha^{(2)} - 12
H^2 \alpha^2 + \frac{3}{4} H^2 |\vec{\nabla} \beta|^2
\nonumber \\
& & +\frac{H}{a} \left( \vec{\nabla}\alpha
\cdot \vec{\nabla} \beta + 2 \alpha \nabla^2 \beta \right) \nonumber \\
& & + \frac{1}{8 a^2} \left( \beta_{,ij} \beta^{,ij} - \left(\nabla^2 
\beta \right)^2 \right ) \,,
\label{G_00} 
\\
\non\\
G^0_{i} &=& G^{0 (0)}_{i} + \delta G^{0 (1)}_{i} + \delta 
G^{0 (2)}_{i} \nonumber \\
&=& - 2 H\alpha_{,i} - 2 H \alpha_{,i}^{(2)} + 8 H \alpha \alpha_{,i}
- \frac{1}{2 a} H \alpha_{,i}\nabla^2 \beta
\nonumber \\
& & + \frac{1}{2 a} \vec{\nabla} \alpha \cdot \vec{\nabla} \beta_{,i}
- \frac{H}{2} \vec{\nabla} \beta_{,i} \cdot \vec{\nabla} \beta - 
\frac{1}{4} \nabla^2 \dot \chi_i^{(2)} \,,
\label{G_0i} 
\\
\non\\
G^i_{j} &=& G^{i (0)}_{j} + \delta G^{i (1)}_{j} + \delta G^{i (2)}_{j}
\nonumber \\
&=& \delta^i_j \left\{ - (3H^2+2\dot{H}) 
+ 2 \alpha (3H^2+2\dot{H})+ 2 H \dot{\alpha} \right.
\nonumber \\ 
& & \left. + \frac{1}{a^2} \nabla^2 \alpha
- \frac{H}{a} \nabla^2 \beta
-\frac{1}{2 a} \nabla^2 \dot{\beta} \right.
\nonumber \\
& & \left. + 2 \alpha^{(2)} (3H^2+2\dot{H})+ 2 H \dot{\alpha}^{(2)}
+ \frac{1}{a^2} \nabla^2 \alpha^{(2)} 
- \frac{H}{a} \nabla^2 \beta^{(2)}
-\frac{1}{2 a} \nabla^2 \dot{\beta}^{(2)}
\right. \nonumber \\
& & \left. + \frac{H}{a} \vec{\nabla}\alpha \cdot \vec{\nabla} \beta 
+\frac{H}{2} \vec{\nabla}\beta \cdot \vec{\nabla} 
\dot{\beta}+\left (\frac{1}{4}
|\vec{\nabla} \beta|^2-4\alpha^2 \right ) \left(3H^2+2\dot{H}\right )- 
\right.
\nonumber \\
& & \left. - 8 H \alpha\dot{\alpha} + \left
(\frac{\dot{\alpha}}{2 a} +2 \frac{H}{a} \alpha\right ) 
\nabla^2 \beta
- \frac{2}{a^2} \alpha \nabla^2 \alpha + \frac{\alpha}{a}
\nabla^2 \dot{\beta} - \frac{1}{a^2} |\vec{\nabla} \alpha|^2 
\right.
\nonumber \\
& & \left.+ 
\frac{1}{8 a^2} \left( \beta^{, \ell m} \beta_{, \ell m} - ( \nabla^2 
\beta )^2 \right) \right\} + \left\{ \frac{1}{2 a} \dot \beta^{, i}_{, j} 
+ \frac{H}{a} \beta^{, i}_{, j} - \frac{1}{a^2} \alpha^{, i}_{, j} \right.
\nonumber \\ & & 
\left. +\frac{1}{2 a} \dot \beta^{(2) \, , i}_{\,\,\,\,\,\,\,\,\,, j}
+ \frac{H}{a} \beta^{(2) \,, i}_{\,\,\,\,\,\,\,\,\,, j} -
\frac{1}{a^2} \alpha^{(2) \,, i}_{\,\,\,\,\,\,\,\,\,, j}
+ \frac{1}{a^2} \alpha^{, i} \alpha_{, j} - \frac{H}{a} \beta^{, i}
\alpha_{, j} + \frac{2}{a^2} \alpha \alpha^{, i}_{, j}
\right.
\nonumber \\ & &
\left.
- \frac{2}{a} H 
\alpha \beta^{, i}_{, j} - \frac{1}{2 a} 
\dot \alpha \beta^{, i}_{, j} - \frac{1}{a}
\alpha \dot{\beta}^{, i}_{, j} + \frac{1}{4 a^2} \left( \nabla^2 
\beta \beta^{, i}_{, j} - \beta^{,i}_{, k } \beta^{,k}_{, j}
\right) \right.
\nonumber \\ & & \left. + \frac{3}{4} H \left (
\dot{\chi}^{(2),i}_{j}+\dot{\chi}^{(2) i}_{\,\,\,\,\,\,\,\, ,j}
+\dot{h}^{(2) i}_{\,\,\,\,\,\,\, j} \right )+ \frac{1}{4}
\left(\ddot{\chi}_{j}^{(2),i}+\ddot{\chi}^{(2) i}_{\,\,\,\,\,\,\,\,
  ,j}+\ddot{h}^{(2) i}_{\,\,\,\,\,\,\, j} \right) \right.
\nonumber \\& & \left. - \frac{1}{4 a^2} \nabla^2 h^{(2) i}_{\,\,\,\,\,\,\, j}
\right\}\,.
\label{G_ii} 
\end{eqnarray}

On the other hand, the axionic coupling between the pseudo-scalar inflaton and the gauge fields changes the energy-momentum tensor, introducing new terms. Up to the second order in perturbation theory, one obtains
\begin{align}
\label{T-tensor}
\non\\
    T^0_{\hspace{0.1cm}0}=&~T^{0\hspace{0.05cm}(0)}_{\hspace{0.1cm}0}+\delta T^{0\hspace{0.05cm}(1)}_{\hspace{0.1cm}0}+\delta T^{0\hspace{0.05cm}(2)}_{\hspace{0.1cm}0} \nonumber \\=& -\qty[\dfrac{\dot{\phi}^2}{2}+V]+\dot{\phi}^2\alpha-\dot{\phi}\dot{\varphi}-V_{\phi}\varphi+\dot{\phi}^2\alpha^{(2)}-\dot{\phi}\dot{\varphi}^{(2)}\nonumber \\&-V_{\phi}\varphi^{(2)}-\dfrac{\dot{\varphi}^2}{2} -\dfrac{\abs{\pmb{\nabla}\varphi}^2}{2a^2}-\dfrac{1}{2}V_{\phi\phi}\varphi^2-2\dot{\phi}^2\alpha^2+2\dot{\phi}\alpha\dot{\varphi}\nonumber \\&+\dfrac{1}{8}\dot{\phi}^2\abs{\pmb{\nabla}\beta}^2-\dfrac{\textbf{E}^2+\textbf{B}^2}{2} \,,
 \non\\
  \non\\
    T^0_{\hspace{0.1cm}i}=&~T^{0\hspace{0.05cm}(0)}_{\hspace{0.1cm}i}+\delta T^{0\hspace{0.05cm}(1)}_{\hspace{0.1cm}i}+\delta T^{0\hspace{0.05cm}(2)}_{\hspace{0.1cm}i}\non\\=&-\dot{\phi}\varphi_{,\hspace{0.05cm}i}-\dot{\phi}\varphi^{(2)}_{,\hspace{0.05cm}i}-\dot{\varphi}\varphi_{,\hspace{0.05cm}i}+2\dot{\phi}\alpha\varphi_{,\hspace{0.05cm}i}+F_{j}^{\hspace{0.1cm}0}F^{j}_{\hspace{0.1cm}i}\,,
 \non\\
 \non\\
 T^i_{\hspace{0.1cm}j}=&~T^{i\hspace{0.05cm}(0)}_{\hspace{0.1cm}j}+\delta T^{i\hspace{0.05cm}(1)}_{\hspace{0.1cm}j}+\delta T^{i\hspace{0.05cm}(2)}_{\hspace{0.1cm}j} \nonumber \\=&\qty(\dfrac{\dot{\phi}^2}{2}-V)\delta^{i}_{\hspace{0.1cm}j}+\qty(\dot{\phi}\dot{\varphi}-\dot{\phi}^2\alpha-V_{\phi}\varphi)\delta^{i}_{\hspace{0.1cm}j}+\nonumber\\&+\Biggl[\dot{\phi}\dot{\varphi}^{(2)}-\dot{\phi}^2\alpha^{(2)}-V_{\phi}\varphi^{(2)}+\dfrac{1}{2}\Biggl(\dot{\varphi}^2-4\dot{\phi}\alpha\dot{\varphi}-\dfrac{1}{4}\dot{\phi}^2\abs{\pmb{\nabla}\beta}^2\nonumber \\&+4\dot{\phi}^2\alpha^2+\dfrac{1}{a}\dot{\phi}\pmb{\nabla}\varphi \cdot \pmb{\nabla}\beta-\dfrac{\abs{\pmb{\nabla}\varphi}^2}{a^2}-V_{\phi\phi}\varphi^2\Biggr)\Biggr]\delta^{i}_{\hspace{0.1cm}j}\nonumber \\&+\dfrac{1}{a^2}\varphi^{,\hspace{0.05cm}i}\varphi_{,\hspace{0.05cm}j}-\dfrac{\dot{\phi}}{2a}\beta^{,\hspace{0.05cm}i}\varphi_{,\hspace{0.05cm}j}-E^iE_j-B^iB_j+\delta^i_{\hspace{0.1cm}j}\dfrac{\textbf{E}^2+\textbf{B}^2}{2} \,.
\end{align}

\end{appendices}

\twocolumngrid